\documentclass[sigconf,screen]{acmart}

\usepackage[inline]{enumitem}
\usepackage{soul}
\usepackage{xcolor}
\usepackage{xspace}
\usepackage{hyperref}
\usepackage{cleveref}
\usepackage{booktabs}
\usepackage{multirow}
\usepackage{balance}
\usepackage{microtype}

\begin{document}

\title[All You Need Is Logs: Improving Code Completion by Learning from Anonymous IDE Usage Logs]{All You Need Is Logs: Improving Code Completion \\ by Learning from Anonymous IDE Usage Logs}

\author{Vitaliy Bibaev}
\affiliation{
    \institution{\textit{JetBrains}}
    \city{Belgrade}
    \country{Serbia}
}
\email{vitaliy.bibaev@jetbrains.com}

\author{Alexey Kalina}
\affiliation{
    \institution{\textit{JetBrains}}
    \city{Munich}
    \country{Germany}
}
\email{alexey.kalina@jetbrains.com}

\author{Vadim Lomshakov}
\affiliation{
    \institution{\textit{JetBrains}}
    \city{Saint Petersburg}
    \country{Russia}
}
\email{vadim.lomshakov@gmail.com}

\author{Yaroslav Golubev}
\affiliation{
    \institution{\textit{JetBrains Research}}
    \city{Belgrade}
    \country{Serbia}
}
\email{yaroslav.golubev@jetbrains.com}

\author{Alexander Bezzubov}
\affiliation{
    \institution{\textit{JetBrains}}
    \city{Amsterdam}
    \country{The Netherlands}
}
\email{alexander.bezzubov@jetbrains.com}

\author{Nikita Povarov}
\affiliation{
    \institution{\textit{JetBrains}}
    \city{Amsterdam}
    \country{The Netherlands}
}
\email{nikita.povarov@jetbrains.com}

\author{Timofey Bryksin}
\affiliation{
    \institution{\textit{JetBrains Research}}
    \city{Limassol}
    \country{Cyprus}
}
\email{timofey.bryksin@jetbrains.com}

\renewcommand{\shortauthors}{V. Bibaev, A. Kalina, V. Lomshakov, Y. Golubev, A. Bezzubov, N. Povarov, T. Bryksin}

\begin{abstract}

In this work, we propose an approach for collecting completion usage logs from the users in an IDE and using them to train a machine learning based model for ranking completion candidates. We developed a set of features that describe completion candidates and their context, and deployed their anonymized collection in the Early Access Program of IntelliJ-based IDEs. We used the logs to collect a dataset of code completions from users, and employed it to train a ranking CatBoost model. Then, we evaluated it in two settings: on a held-out set of the collected completions and in a separate A/B test on two different groups of users in the IDE. Our evaluation shows that using a simple ranking model trained on the past user behavior logs significantly improved code completion experience. Compared to the default heuristics-based ranking, our model demonstrated a decrease in the number of typing actions necessary to perform the completion in the IDE from 2.073 to 1.832.

The approach adheres to privacy requirements and legal constraints, since it does not require collecting personal information, performing all the necessary anonymization on the client's side. Importantly, it can be improved continuously: implementing new features, collecting new data, and evaluating new models --- this way, we have been using it in production since the end of 2020.

\end{abstract}



\maketitle

\vspace{0.3cm} 

\section{Introduction}

Integrated development environments (IDEs) are among the most important tools used in writing software code~\cite{hou2009empirical}. They provide developers with various smart features that increase their productivity: highlighting of code, syntactic checks, various automatic quick-fixes~\cite{mucslu2012speculative}, automatic refactorings~\cite{lee2013drag, oo2018dynamic}, and others.

To evaluate how various features are being used, it is possible to collect \textit{logs} from IDEs. This can be used for high-level analysis of user workflows~\cite{ardimento2019mining, ioannou2018mining, amann2016study} or for evaluating the usefulness of a particular feature, for example, automatic refactorings~\cite{murphy2011we, negara2013comparative}. However, one needs to be careful when dealing with usage logs, similar to any user data. Researchers often use the data from a limited group of volunteers with explicit consent~\cite{snipes2015practical}, while processing the data of thousands of users in the wild requires stronger protection of their privacy in accordance with various regulations, such as the General Data Protection Regulation (GDPR)~\cite{gdpr}.

\begin{figure}[h]
    \centering
    \includegraphics[width=\columnwidth]{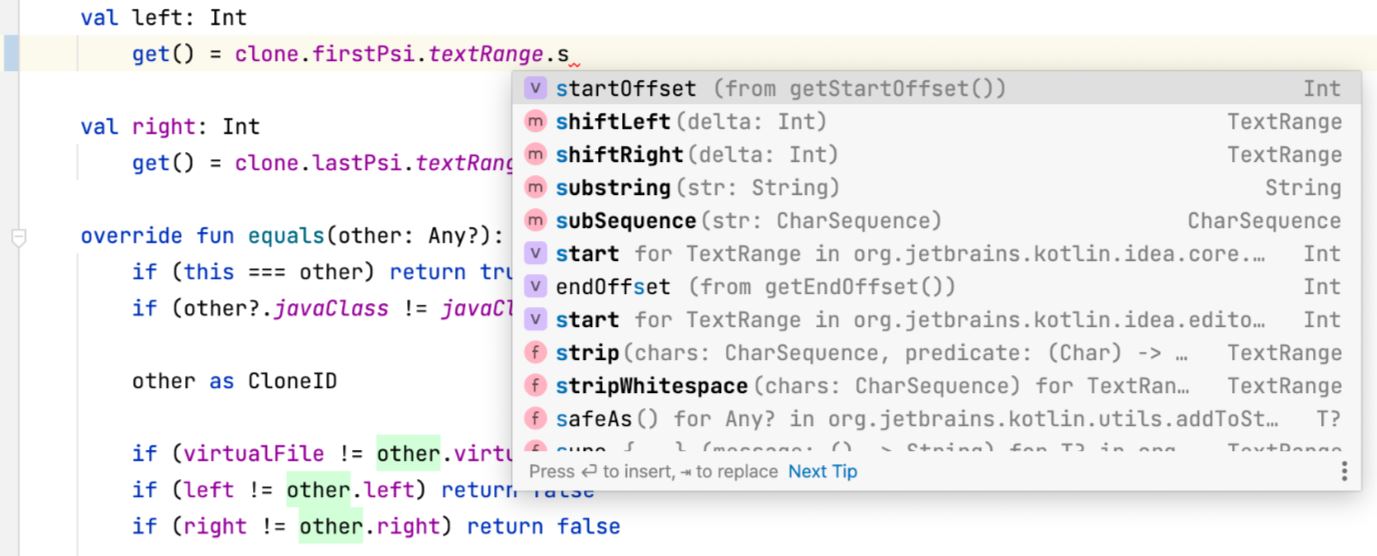}
    \caption{An example of code completion in the IDE.}
    \label{fig:example}
    \vspace{0.2cm}
\end{figure}

One of the key features that defines an IDE is \textit{code completion}~\cite{svyatkovskiy2020intellicode, svyatkovskiy2019pythia, izadi2022codefill, ciniselli2022extent, amann2016study}. Code completion speeds up the programming process by automatically suggesting code that the developer is about to write, while also helping them avoid possible typos. An example of code completion in action is presented in Figure~\ref{fig:example}.

IDEs based on the IntelliJ Platform~\cite{kurbatova2021intellij} (IntelliJ IDEA, PyCharm, WebStorm, etc.), developed by JetBrains~\cite{jetbrains}, naturally also have a code completion feature. The default implementation uses a static analyzer to generate the candidates by parsing syntactic and semantic models of the project and collecting the information about the used entities, as well as the given position of the caret. Then, a heuristics-based ranking is used to sort the suggestions before presenting them to the user. Despite the good quality and fast inference, this leads to three major problems:

\begin{enumerate}[label=(\arabic*)]
    \item the heuristics are based on certain assumptions that can be hard to verify quantitatively and may be statistically wrong;
    \item over time, maintaining a large number of heuristics becomes cumbersome and it gets harder to add new ones without breaking the existing ones;
    \item supporting a new language for completion is difficult, even when a lot of heuristics might be reused.
\end{enumerate}

To overcome these problems and further improve code completion, an existing heuristics-based system can be augmented by adding a machine learning (ML) based ranking as the final step~\cite{Bruch2009, Nguyen2013, Raychev2014, Proksch2015inteli}. Training an ML ranking model that will rearrange suggestions from the static analyzer requires ground truth data, \textit{i.e.}, a labeled list of code completion suggestions. Existing approaches can be divided into two principal categories from the standpoint of the data that they use for training:
\begin{enumerate*}[label=(\arabic*)]
    \item \textit{synthetic}: that analyze the existing body of source code~\cite{Nguyen2013, Hindle2012, Raychev2014}, and
    \item \textit{real user behaviour}: that employ logs, source code, and edits from real users~\cite{Robbes2008, Aye2020real}.
\end{enumerate*}
In a recent paper, Hellendoorn et al.~\cite{hellendoorn2019code} showed that synthetic data significantly differs from the real-world usage, so training on it may cause worse performance for the end-users. For this reason, utilizing the usage logs seems like a promising solution.

In this work, we propose and evaluate an approach that allows to use the information from the users without violating their privacy or collecting any personal information. We propose to decouple the extraction of data by designing a set of features that are computed on the side of the client and are then collected anonymously from logs. We then use this data to train a model for ranking the candidates for code completion and compare its performance with the default heuristics-based approach.

Specifically, we defined a number of different features that describe the prefix before the caret, the context around the suggestions, the entities defined and used before the caret, the history of selecting any given suggestion by the user, 
etc. Then, we integrated the collection of these features into the Early Access Program (EAP)~\cite{eap} of IntelliJ-based IDEs and collected anonymized logs from the users who agreed to send them to us. The logs include the values of the defined features for each suggestion, as well as the label: positive for the suggestion that was actually selected by the user, and negative for all the rest. This way, no sensitive information is collected (\textit{e.g.}, the user's code), but the ranking model can still be trained. We used a CatBoost model~\cite{prokhorenkova2019catboost}, because it is fast, lightweight, and can be easily converted for the JVM.

We carried out two different evaluations of the described approach using the data from Python-based projects in PyCharm: an \textit{offline} evaluation and an \textit{online} evaluation. For the offline evaluation, we collected two datasets of logs from two different groups of users: the first week contained 54,540 completion sessions and was used for training the model, the second one contained 72,131 sessions and was held out for testing. In this setting, the proposed ML-based ranking demonstrated the Recall at Top-1 (\textit{i.e.}, cases when the correct suggestion is at the very top of the list) of 0.870 against 0.761 for default heuristics-based ranking. 

However, while such an evaluation is much better than a synthetic one, even it does not tell the full story. To see the results in action, we also performed an online evaluation that constituted an A/B test: some users were given the base code completion using heuristics, and some were given suggestions ranked using our model. Such a setting allowed us to use more practical, intuitive metrics of the completion quality. Among other things, this evaluation showed that the average number of typing actions required to finish the completion lowered from 2.073 for the heuristics-based ranking to 1.832 for the ML-based one, indicating the real saving of time for the users. The latency increased slightly from 92.3 ms to 119.3 ms, thus remaining in the comfortable range. 

One important aspect of using the EAP versions of an IDE is that this allows us to repeat the described process cyclically, in the new releases. This way, new data can be obtained from the users in a constant stream, and, importantly, new models can be evaluated and compared in live A/B tests, thus ensuring constant evolution of the quality. This process takes place in EAP versions of IntelliJ-based IDEs since the end of 2020.

Overall, the paper presents the following contributions: 

\begin{enumerate}
    \item \textbf{Approach} for enhancing the quality of code completion in real-world scenarios that:
    \begin{itemize}[leftmargin=0.45cm]
        \item[---] formulates code completion as a ranking problem;
        \item[---] consists of a feature-based CatBoost model trained on real anonymized user behavior data collected without violating their privacy;
        \item[---] is language-agnostic: while some specific features and specific trained models can be used only for a specific language, the approach as a whole, as well as a lot of general features, does not depend on the language;
        \item[---] meets the requirements of the real-world industrial application: always produces syntactically correct code, results in a relatively small model (less than 500 KB) with low inference time (20-30 ms);
        \item[---] can be continuously employed in cycles of gathering data and comparing models in live A/B tests.
    \end{itemize}

    \item \textbf{Evaluation} of the proposed approach in two different settings that employ real user data from PyCharm:
    \begin{itemize}[leftmargin=0.45cm]
        \item[---] offline evaluation on the held-out user data, which demonstrates that ML-based ranking shows the Recall at Top-1 of 0.870 over the heuristics-based ranking that has the Recall at Top-1 of 0.761;
        \item[---] online evaluation via an A/B test in the IDE, which shows that the average number of typing actions necessary to perform completion lowers from 2.073 for the heuristics-based ranking to 1.832 for the ML-based one.
    
    \end{itemize}
    \item \textbf{Insights} into the results of using the proposed approach in a real-world industrial IDE and dealing with various constraints that such a setting implies. Models obtained using the described approach are currently being employed in almost all major IntelliJ-based IDEs.
\end{enumerate}

The rest of the paper is organized as follows. In \Cref{sec:background}, we briefly discuss the problem of code completion and the constraints it faces in the real-world applications. \Cref{sec:approach} describes in detail the proposed approach to collecting the data and building the model, while \Cref{sec:evaluation} describes two different evaluations that we conducted. In \Cref{sec:discussion}, we share insights about the usage of the proposed approach and discuss open challenges in the field. \Cref{sec:related} describes the existing related works, \Cref{sec:threats} discusses the threats to the validly of our work, and, finally, \Cref{sec:conclusion} concludes the paper and discusses possible directions for future work.
\section{Background}
\label{sec:background}

\begin{figure*}[h!]
    \centering
    \includegraphics[width=0.9\textwidth]{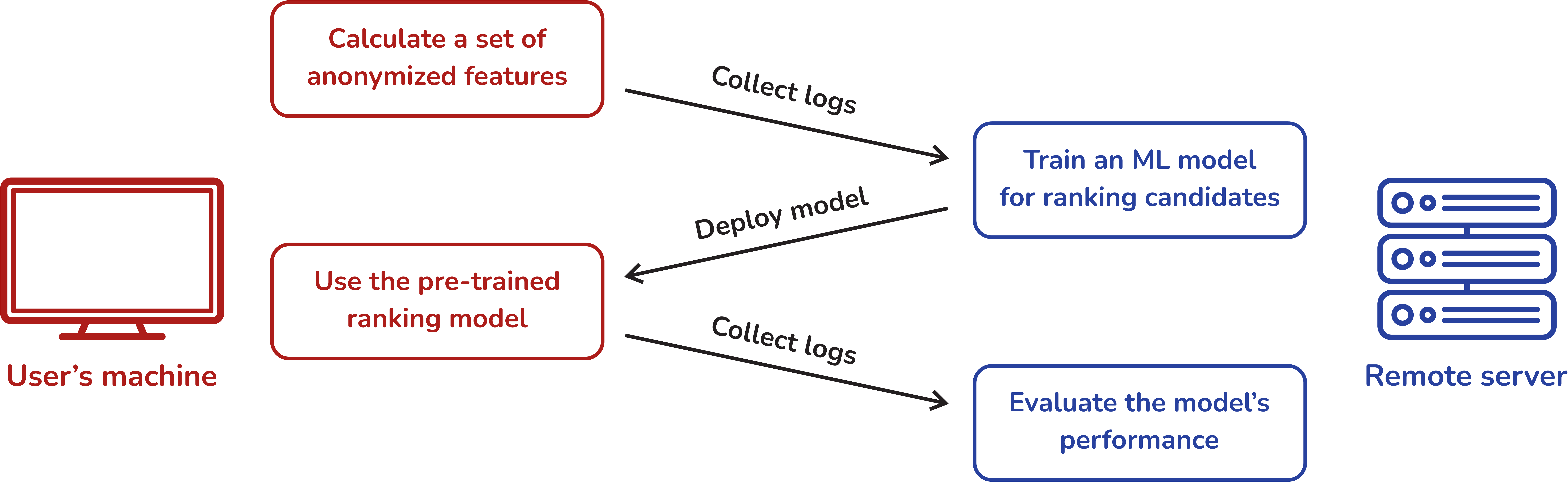}
    \caption{The pipeline of the proposed approach.}
    \label{fig:approach}
\end{figure*}

The idea of code completion as an IDE feature is to suggest code to the user before they can manually type it, thus saving them the time and effort. Additionally, code completion can help in project exploration, providing new users with an opportunity to see entities from different parts of the code base. Since there is a lot of variance in the process of writing code, the IDE usually provides a ranked list of possible completions for the user to choose from.

Code completion systems vary by the granularity of their suggestions: from tokens and API usages to lines or even entire methods. While early research focused mostly on narrow API-level completion~\cite{Bruch2009, pletcher2009bcc, Hou2011}, modern language models based on neural networks vary from fine-grained, using every possible lexical token type (delimiters, operators, white spaces, keywords, etc.)~\cite{Hindle2012}, to coarse-grained, predicting entire lines of code~\cite{wang2020towards, chen2021evaluating}. 

In this work, we specifically focus on token-level code completion as the most balanced and important option: it is more practical than heavy full-line code completion and more general than just API recommendation. In practice, over time, effective token-level code completion can save the users a lot of effort. However, our approach is easy to extend to other types of completion, and we leave applying the usage of logs for the full-line version of code completion~\cite{wang2020towards} for subsequent work.

The set of possible candidate suggestions in code completion is determined by the implementation of the \textit{candidate provider}~\cite{svyatkovskiy2021fast}. Code completion systems in IDEs rely on \textit{static analysis} to retrieve the list of possible candidate suggestions, since it is vital for suggestions to be syntactically correct. The list of candidates is generated by considering the position of the caret, the grammar of the programming language, code entities that are defined and used in the opened file, other files in the project, and the used libraries.

Before showing the list of candidates to a user, the IDEs \textit{rank} them, typically employing a series of hard-coded heuristics. These heuristics range from trivial and language-agnostic, \textit{e.g.}, which entities were used the most in the opened file, to specific and language-dependent ones, \textit{e.g.}, type matching. 

Thus, there are at least two different ways to improve token-based code completion: (1) change the set of possible candidate token suggestions, or (2) change the order in which these suggestions are presented to the user. In this work, we focus on the second one, building on a number of works that enhance the quality of ranking by employing statistical and machine learning approaches~\cite{svyatkovskiy2021fast, Proksch2016eval, Proksch2015inteli, Aye2020real}. It is vital to focus on improving the ranking of the suggestions, since in general, modern static analyzers already generate candidates very well, and their ``recall'' is rather good, so it becomes a matter of prioritizing syntactically correct suggestions. The general idea of using machine learning for this task consists in choosing the necessary \textit{features} that describe the context around the completion instance and the candidates, and collecting \textit{data} for training the ranking model. Since the model is trained to be used in an industrial setting, inside the user's IDE, this imposes a set of important constraints on the entire process.

Firstly, because of the slow and energy-consuming nature of the model training, it must happen on a remote server, and the user should simply receive a pre-trained model as a part of their IDE. Secondly, the inference of the model has to take place on the user's machine, be fast and reliable. This includes not using any special hardware (\textit{i.e.}, GPU) and not accessing the network. The last constraint is important for several reasons:

\begin{itemize}
    \item[---] Inferencing the model over network introduces additional latency to the process.
    \item[---] Developers might work without access to the internet.
    \item[---] Developers may not be comfortable with sending any information to a remote server.
\end{itemize}

The concern about the privacy of the user's data is crucial in this area, and also directly relates to the data collection. As was mentioned above, training the necessary model requires having the labeled data of code completion sessions. Some of the existing works use \textit{synthetic} data and analyze the existing body of source code~\cite{Nguyen2013, Hindle2012, Raychev2014}. However, Hellendoorn et al.~\cite{hellendoorn2019code} demonstrated that in the case of code completion, synthetic data may differ significantly from the real-world data, so such data collection pipelines are inherently flawed. For this reason, other works analyze \textit{real user behaviour}: they gather the data from logs, source code, and edits made by real users~\cite{Robbes2008, Aye2020real}. In such a setting, privacy requirements are also critical and must be taken into account.

In short, the privacy requirements regulate the gathering of information that might identify the user. While this obviously includes any personal information like the name or the e-mail, in certain legislatures, this can also include the code itself. In particular, in recent years, a lot of laws were passed such as EU's General Data Protection Regulation (GDPR)~\cite{gdpr} or California Consumer Privacy Act (CCPA)~\cite{ccpa} that carefully regulate users' privacy and data collection. User data should be collected only in anonymized, depersonalized form, without their code or any code metric that is too revealing of their identity. This is crucial both in the sense that it incentivizes inferencing the model on the user's machine, without sending any data, and also impacts the possible ways of collecting user data for training.

In this work, we propose, describe, and evaluate improving the performance of code completion by collecting anonymized logs of real completion sessions and training an ML model based on them.
\section{Approach}
\label{sec:approach}

Employing user logs to improve code completion requires finding a proper way to collect and utilize these logs. Taking into account the mentioned limitations, we propose the following pipeline for collecting the data and training the model, shown in Figure~\ref{fig:approach}:

\begin{enumerate}
    \item A set of features is calculated on the user's machine during their completion sessions.
    \item This data is anonymized and collected as logs without any identifying personal information.
    \item A model is trained and evaluated on the server using a large amount of such anonymized data.
    \item This pre-trained model is deployed to future users, from whom we can collect new logs to evaluate the model.
\end{enumerate}

In such a setting, all the requirements are met: no sensitive personal data is collected (neither the suggestions nor the surrounding code), the resource-expensive training of the model happens on the server, and the usage of the model requires nothing from the user. What is even more important, however, is that the described sequence represents just one iteration of the process: in parallel to evaluating a model, new data can be collected from other users, new models can be trained, etc. This opens up the possibility to compare the models in a real industrial setting. The implementation of the described pipeline requires three key things:

\begin{enumerate}
    \item Designing a set of features to be collected and a data format for storing and utilizing them.
    \item Selecting a machine learning model to train on this data that could be conveniently inferenced in the JVM-based IDE.
    \item Creating a setting for continuously collecting data and testing models in the form of A/B tests using the Early Access version of IntelliJ-based IDEs.
\end{enumerate}

Let us now describe each of these three items in greater detail.

\subsection{Data Collection}

As was previously stated, we cannot train the necessary model on the user's machine or send personal information about them (including the code) because of privacy concerns. Because of this, the data collection consists in (1) designing a set of anonymous features that would describe the completion session, and (2) sending this data to the server and transforming it for training the model.

It should be noted right away that while a lot of modern works argue in favor of end-to-end neural models instead of manual feature extraction~\cite{svyatkovskiy2021fast}, in our setting feature extraction represents a benefit: it provides a clear organizational way to control and manually audit what exactly gets collected and sent, via code reviews.

The collected data consists of \textit{completion sessions}, \textit{i.e.}, cases when the pop-up window with suggestions appeared. Each completion session, in its turn, contains one or several \textit{look-ups}, \textit{i.e.}, specific lists of suggestions. One session can have several look-ups if the user typed additional characters when the pop-up already appeared, thus changing the context and filtering the list. Finally, each look-up consists of specific \textit{suggestions}, \textit{i.e.}, individual tokens that are ranked for the user to choose from.

\subsubsection{Feature Selection}
\label{sec:features}

The aim of the features is to describe the completion context of any given suggestion as best as possible without actually giving away any of the written code. In this case, the \textit{context} relates not only to the code right before the caret, but anything that can influence the choice: the information about the suggestion itself, the history of choosing, etc. In our experiments, we evaluated many different features and their importance for the quality of the model. It is very important to note that the proposed approach is inherently language-agnostic, so some basic features are also the same for different languages, while some features remain language-specific. 

There are several main groups of language-agnostic features:

\begin{itemize}
    \item[---] \textbf{Information about the prefix.} A major source of information about the completion is the \textit{prefix}, \textit{i.e.}, the already typed part of the token, right before the caret. The information includes the length of the prefix, the number of matched characters between the prefix and the suggestion, whether the match is case sensitive, whether it is exact, etc.

    \item[---] \textbf{Syntactic context.} The completion session does not happen in a vacuum, it takes place in a certain code location, in a certain part of the codebase. All of this can influence the necessary token. Thus, features that describe the syntax include whether the suggestion is a language keyword, whether the suggestion is the element from the same file or module, whether it is from a third-party library, etc.

    \item[---] \textbf{Syntactic history.} Additionally, and crucially, it is very important to catch the dynamic nature of software development and not treat the opened file as a static object. Previous actions of the user may indicate their intent better than the basic proximity of objects in the code. As such, we analyze the prior navigation of the user to see whether they navigated to the definition of the suggestion before, etc.

    \item[---] \textbf{Session history.} Also, some features describe the temporal aspect of completion sessions: the duration of a given session in seconds, whether the user has already selected the given suggestion before, etc.
\end{itemize}
 
\textbf{Language-specific} features include the context of the suggestion (\texttt{if} block, \texttt{for} block, etc.), the type of the suggestion (method, class), the usage frequency, or something even more specific like the number of lines between the caret and the \texttt{\_\_init\_\_} method of the enclosing class in Python.

Initially, there were several hundred different features for each language. Feature selection consisted of two stages. On the first stage, we filtered out all the features that were not used by the model during training. On the second stage, we calculated permutation importance~\cite{breiman2001random} for the features and filtered out those that were not useful. All the remaining features were used in the final production model. We evaluated our approach on different languages: for example, for the Java language, the total final number of the used features was 146, for the Python language, it was 188. 

\vspace{0.05cm}

\subsubsection{Data Format}
\label{sec:data_format}

 \begin{figure*}[h!]
    \centering
    \includegraphics[width=0.9\textwidth]{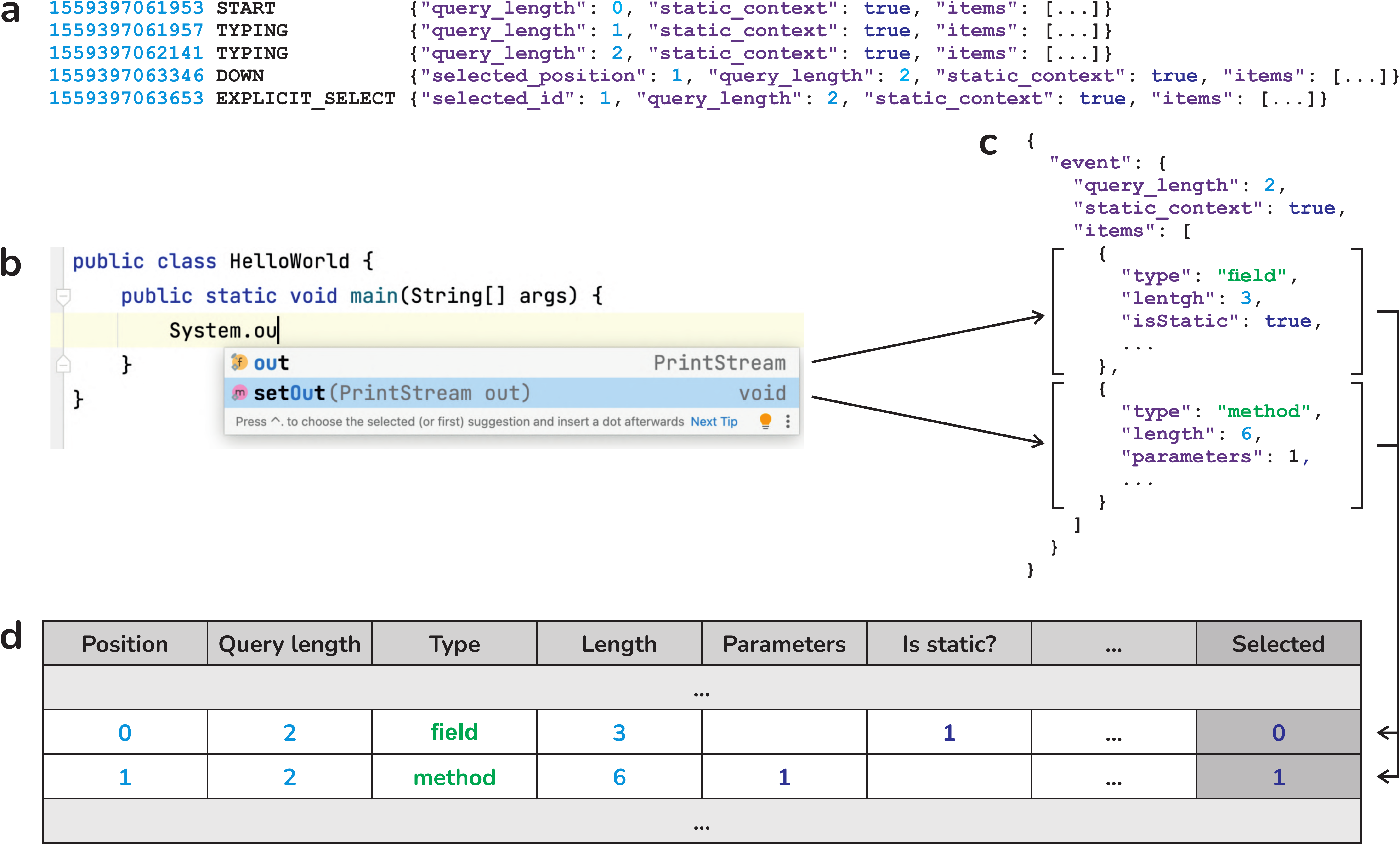}
    \caption{Data collection and data format. In this example, the user typed two characters of the token and selected the second suggestion. (a) The simplified general form of the collected logs during actions, (b) the view inside the IDE right before selection, (c) the JSON structure of the collected data, including several simple features, (d) the final representation of the data with the target column that the model trains on.}
    \label{fig:data}
    \vspace{-0.4cm} 
\end{figure*}

Once the features are chosen, they are deployed for the collection of logs into the Early Access version of the necessary IDE. The collection occurs for each \textit{action} in each completion session. The possible actions are: the completion session \textit{started} (\textit{i.e.}, the pop-up appeared); the user \textit{typed} an additional symbol (thus adding another loop-up to the session, changing the prefix, the context, and filtering the list of suggestions); the user \textit{went up or down} the list of suggestions using arrow keys; or the completion session \textit{ended} (in one way or another). The collected data contains the following information: 

\begin{itemize}
    \item[---] User ID, random and anonymous.
    \item[---] Timestamp of the event.
    \item[---] The way the completion session started: \textit{manually}, by pressing the specific hot-key combination, or \textit{automatically}, after the user started typing.
    \item[---] Project-, file-, or snippet-level features that are the same for all suggestions in the look-up.
    \item[---] The list of individual suggestions and their features.
    \item[---] The way the session ended and the selected option, if any.
\end{itemize}

There are four possible ways for the completion session to end: 
\begin{itemize}
    \item[---] \textbf{Explicit select}, if the user selected a suggestion and it was successfully auto-completed.
    \item[---] \textbf{Typed select}, if the user completely manually typed a suggestion that was in the list.
    \item[---] \textbf{Explicit cancel}, if the user explicitly canceled the completion (\textit{e.g.}, by pressing ESC or reverting).
    \item[---] \textbf{Typed cancel}, if the user completely manually typed a token that was not in the list, thus also ending the session.
\end{itemize}

A simplified example of the collected logs is presented in Figure~\ref{fig:data}. In this example, the full logs (Figure~\ref{fig:data}a) show that the completion session started, then two characters were typed, after which the user pressed the down arrow once and selected the token, thus successfully finishing the completion. For each action, the full list of features is recorded. Figure~\ref{fig:data}b and Figure~\ref{fig:data}c show the suggestions that were present in this example before selection, and several basic features calculated for them.

For training the model, we use data points that ended with \textit{Explicit select} or \textit{Typed select}, \textit{i.e.}, where there is ground truth. After the user selects a certain suggestion, it is labeled as the correct option in all the look-ups in this completion session. For example, if the user selected the token after typing two characters, the suggestion was already in the list after the first character, perhaps, lower, and can still be considered the correct option in that list. Figure~\ref{fig:data}d shows that the selected suggestion (``\texttt{setOut}'') was labeled with 1 as the target value, while the other suggestion was labeled with 0.

The remaining data can be used to evaluate the quality of the used completion. The timestamps can be used to calculate the time that the user spends evaluating their options, the typing actions can be used to see how many are required on average to finalize the completion, and we can also measure how many sessions ended in each of the four possible ways. These different metrics will be described in greater detail in Section~\ref{sec:evaluation}.

\vspace{-0.2cm} 

\subsection{Model}
\label{sec:model}

Similar to the other works motivated by industrial application~\cite{Aye2020seq, svyatkovskiy2021fast}, we focus on improving the ranking of the candidates and formulate this task as a \textit{Learning to Rank} problem~\cite{li2011short}. Researchers explored a similar architectural approach~\cite{Bruch2009, Proksch2015inteli}, referred to as \textit{structural feature selection}~\cite{hellendoorn2019code} and \textit{feature-based models}~\cite{svyatkovskiy2021fast}, only for the limited context of API recommendations over the closed vocabulary of possible API calls. 

In Section~\ref{sec:background}, aside from the limitations that relate to privacy, we also mentioned performance-related limitations. The model must not rely on any specific hardware and must be lightweight to operate on the user's machine. More specifically, the size of the model should not be more than 2 MB and the inference must take tens of milliseconds in order not to make the completion generation too noticeable for the user. Such strict limitations come from the fact that during the completion session, the model will be inferenced after each new character. Last but not least, the model must be easy to integrate into the JVM-based architecture of the IDE.

To meet these requirements, in particular the last one, we decided to use a model based on the \textit{Decision Tree} algorithm. This architecture is a good choice for us, because it is easy to convert into the \texttt{if-else} code representation, and thus it can be used inside the JVM process. Specifically, we used a CatBoost~\cite{prokhorenkova2019catboost} model with a built-in QuerySoftMax loss function~\cite{softmax, cao2007learning}.

We experimented with a number of alternative models, including lightGBM~\cite{ke2017lightgbm}, as well as feed-forward and Transformer-based neural models. None of these options came close to the performance and the operational simplicity of the tree-based model. Given the severe cap on the size of the model, it is difficult to train a well-performing Transformer, and no other architectures demonstrated results better than CatBoost. In the end, the CatBoost model was the only one actually deployed to production and thus will be the focus of the rest of the paper.

As discussed in Section~\ref{sec:data_format} and shown in Figure~\ref{fig:data}d, an individual data point consists of a list of several code completion suggestions, the accepted one being a positive example and the rejected ones being negative. It is also important to understand that the users mostly care about what is on the very top of the list. For this reason, we needed to choose a loss function that would prioritize the correct suggestion being at the top of the list. CatBoost comes with such a loss function, QuerySoftMax~\cite{softmax}, described previously in the work of Cao et al.~\cite{cao2007learning}. In this setting, given a list of suggestions, the model learns to boost the correct one to the top.

The model was trained using the native CatBoost framework, then converted into the lightweight representation of a series of \texttt{if-else} operators. The final model was implemented and inferenced in the Java language.

\subsection{Continuous Collection and Validation}
\label{sec:continuous}

\begin{figure*}[h]
    \centering
    \includegraphics[width=0.9\textwidth]{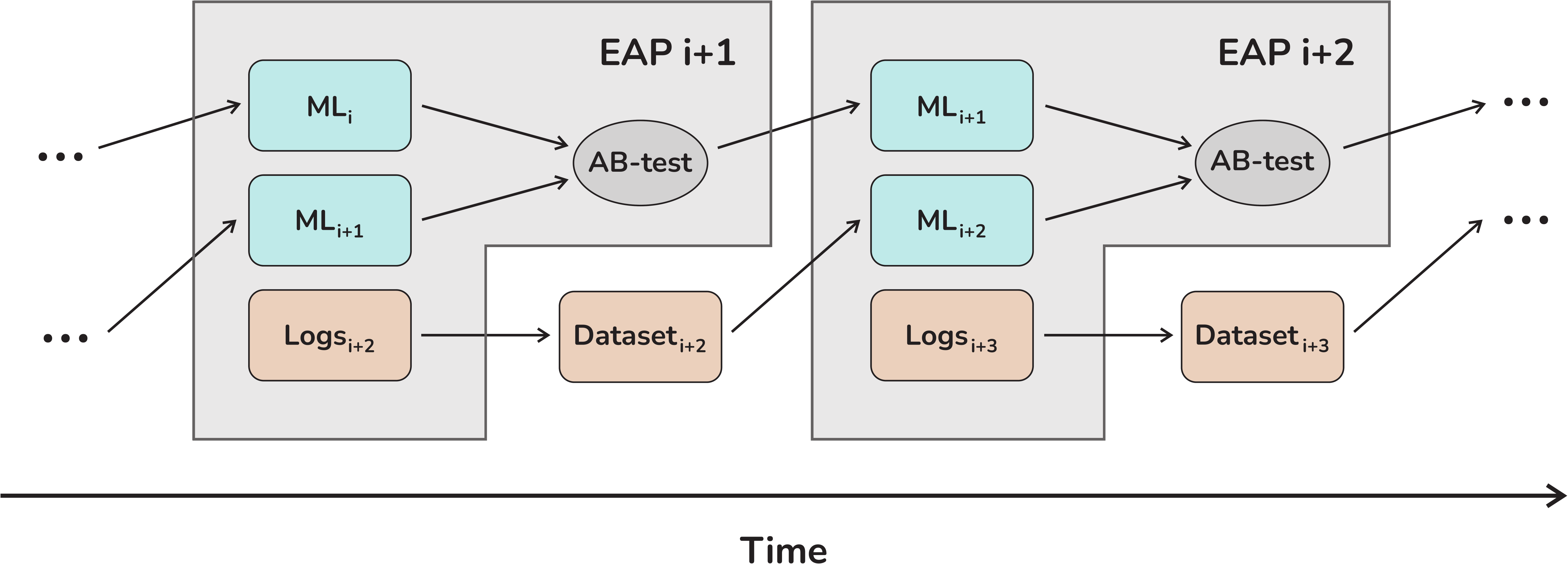}
    \caption{The continuous way of collecting the data and comparing the models in the Early Access versions of the IDEs. In every EAP, one group of users gets the current best model and another gets the new tested model, between which an A/B test is run. At the same time, from a separate groups of users new logs are being collected that are then used to build a new dataset and train the next iteration of the model.}
    \vspace{-0.4cm}
    \label{fig:data_collection}
\end{figure*}

One of the most important features of the proposed approach is that the gathering of new data and the evaluation of the models can occur in parallel, since both processes rely on collecting logs. At the same time, the cyclical nature of IDE releases means that this process can be repeated over and over, always collecting new data (with new features), training fresh models, and evaluating them on real users against the current best.

This process relies on the Early Access Program (EAP)~\cite{eap} provided for IntelliJ-based IDEs. It works as follows. In the two-month period before each release of an IntelliJ-based IDE, the EAP takes place. If a user agrees to participate in the EAP, they are given free access to a beta-version of the upcoming commercial release, in which they can evaluate and try out new features. In exchange, they agree to provide anonymous logs of their experience that the developers of IDEs can use to fix any existing bugs. In this setting, it is not only possible to collect data about code completion or evaluate a single model, but also carry out a fully-fledged A/B test in the process. The idea of this cycle is demonstrated in Figure~\ref{fig:data_collection}. Let us now describe it further.

Let us say that in the EAP of the version $i$, a certain model $ML_{i}$ was working and some data was collected. Then, we take this data and train a new model that we think will do better, let us call it $ML_{i+1}$. At the same time, perhaps, we already developed some new features that we want to evaluate. So, when the EAP version $i+1$ comes around, we implement the collection of the new features, and then divide the users of the EAP into three groups:

\begin{enumerate}
    \item The $\mathbf{ML_{i}}$ group receives the $ML_{i}$ model that was obtained in the previous version, the current best.
    \item The $\mathbf{ML_{i+1}}$ group receives the $ML_{i+1}$ model, the new one.
    \item Finally, the separate \textbf{Logs} group also receives the previous \textbf{$ML_{i}$} model and is used to collect data for the next cycle.
\end{enumerate}

From all three groups, the logs are collected. This way, an A/B test is being conducted between the two ML-based models, and with the help of the metrics calculated from logs and statistical tests, we can discover the best of them. Meanwhile, the logs collected from the third group can be used to create a training dataset, on which we can train a new model, $ML_{i+2}$. Then, during the EAP version $i+2$, we can repeat the process, comparing models $ML_{i+1}$ and $ML_{i+2}$, collecting new data with the new features, and so on. 

It is also possible to compare more than two models at once and fit several iterations into a single EAP cycle. In this sense, the EAP is very convenient, because it is experimental by definition, and so the users are expecting to receive frequent updates that change the behavior of the IDE --- something that would be unacceptable in a major release.
The described process has been running in the EAPs of major IntelliJ-based IDEs since the end of 2020. This allowed us to evaluate different models and a lot of different features. 
\section{Evaluation}
\label{sec:evaluation}

To highlight the usefulness of the proposed approach for using logs, as well as the ability of the CatBoost-based model to improve the quality of code completion, in this section, we describe two different evaluations that we conducted to compare it with the baseline heuristic-based ranking.

\subsection{Data}
\label{sec:eval_data}

The data for the experiments was collected in September and October of 2020 from Python-based projects in PyCharm. Overall, we collected two sets of data in two consecutive weeks. The first set contained 54,540 completion sessions from 2,623 unique users. This set was split into two in the ratio of 80\%/20\% of individual users: one was used for training the model, and the other was used for tuning hyper-parameters. The other set, collected during the second week, contained 72,131 completion sessions from 2,086 unique users, different from the ones in the first week. This set was held out for testing. This way, the testing data was located later in time from the training data and came from different users. The model utilized 188 Python-based features described in Section~\ref{sec:features}.

\subsection{Offline Evaluation}

\subsubsection{Methodology}

The first evaluation was carried out \textit{offline}, meaning that after the data from Section~\ref{sec:eval_data} was already collected, the model was trained on the data from the first week and then tested on the held-out data from the second week. On the same held-out data, the heuristics-based ranking was also evaluated.

To evaluate the performance of the models, we used the Recall at K metric ($\mathbf{R@K}$). This metric represents how often the correct answer was in the top $K$ suggestions of the look-up. Naturally, $\mathbf{R@1}$ is especially important, since it tracks how often the correct answer was at the very top, convenient for the user.

Additionally, besides calculating the metric for \textit{all look-ups} in all completion sessions, we also separately calculate them only for the \textit{initial look-ups}, meaning the first look-ups in each session, before typing any additional characters. This metric correlates well with the user experience, since it is very important to suggest the correct item from the very start --- if the user has to type additional characters, they are unlikely to pause and analyze the list after every keystroke, at this point, they are more likely to simply type the token themselves.

\subsubsection{Results}

\begin{table}[t]
    \centering
    
    \caption{The results of the offline evaluation. $\mathbf{R@K}$ stands for Recall at K, $\mathbf{init}$ stands for initial look-ups.}
    \vspace{-0.2cm}
    \begin{tabular}{ccccc}
        \toprule
        \textbf{CC System}     & $\mathbf{R@1_{all}}$ & $\mathbf{R@5_{all}}$& $\mathbf{R@1_{init}}$ & $\mathbf{R@5_{init}}$ \\
        \midrule
        \textbf{Heuristics} & 0.761           & 0.957 & 0.634 &     0.918    \\
        
        \textbf{CatBoost}    & \textbf{0.870}           & \textbf{0.981} & \textbf{0.799} & \textbf{0.959}  \\
        \bottomrule
    \end{tabular}
    \vspace{-0.4cm}
    \label{tab:evalOffline}
\end{table}

Table~\ref{tab:evalOffline} summarizes the comparison between the default ranking and our model. It can be seen that our model provides better results for all the evaluated metrics. The correct token is more often shown in the first position, and is higher located overall, with both $\mathbf{R@5}$ metrics reaching very high values. The largest increase can be seen for the most important metric --- $\mathbf{R@1_{init}}$, or the ratio of correct suggestions at the very top of the list during the initial look-ups. The CatBoost model increased this metric by 16.5 percentage points, from 0.634 to 0.799, indicating a much better performance for the user in the most crucial moments. 

It is interesting to note that different types of tokens demonstrated different increases in recall. For example, functions from the global scope (\textit{e.g.}, \texttt{abs()}) demonstrated a significant increase in quality. There are a lot of such functions in Python that are difficult to complete using heuristics due to the lack of type checking, whereas an ML model learned how they are used. An opposite example, where the increase in recall was small, is API calls. The reasons for this are that, firstly, the default completion already works well in these cases, and secondly, it is often difficult to understand what API method to use from the local context. A promising direction for such cases is to learn from other API usages~\cite{Bruch2009}.

\subsection{Online Evaluation (A/B Test)}
\label{sec:online}

\subsubsection{Methodology}

The offline evaluation demonstrated good results, but it is very important to understand its limitations. While these results represent realistic usage scenarios~\cite{hellendoorn2019code}, the offline metrics themselves are only proxy-metrics for real \textit{usability} improvements: having the correct result higher in the list indicates more efficient work, but does not directly show it.

To explicitly measure more interpretable, human-oriented metrics, we conducted an online evaluation in the form of an A/B test. The implementation of the A/B test was the same as described in Section~\ref{sec:continuous}, only with the evaluated models being our CatBoost model and the default heuristics-based ranking. In total, 231 users were allocated to the heuristics-based Control group and 246 users --- to the CatBoost group. From the first group, 66,144 completion sessions were collected, from the second --- 74,346 sessions.

The setting of an A/B test allows us to use all the collected data described in Section~\ref{sec:data_format} to further measure the performance, using several dozen specific metrics that describe every aspect of the completion sessions. We will report several key ones:

\begin{itemize}
    \item[---] \textbf{Explicit select}. Firstly, we can simply measure the fraction of the sessions that ended in the explicit selection of the token, the best possible result.
    \item[---] \textbf{Typed select}. In contrast, we can measure the fraction of sessions that ended with typed select, meaning that the user typed the token themselves, even though it was in the list.
    \item[---] \textbf{Typing actions}. To measure the actual increase in the users' productivity, we can calculate the average number of typing actions (typed characters) in the completion session. For this metric, we used a cut-off at 0.99 percentile to remove anomalous completion sessions with an extremely large number of typed characters.
    \item[---] \textbf{Prefix length}. To specifically gauge the effectiveness of \textit{explicit} select, we can also compare the average length of prefix at the moment of explicit selection. This will also demonstrate when it is necessary to type fewer characters before explicitly finishing the session.
    \item[---] \textbf{Manual start}. Finally, to measure the overall reliance of users on code completion, we can measure the fraction of sessions that were started manually, \textit{i.e.}, using the hot-key combination.
\end{itemize}

To test the statistical significance of the difference between models, we employed bootstrap~\cite{Efron1979BootstrapMA}. When comparing the models, we form 1,000 bootstrapping re-samplings of completion sessions \linebreak grouped by users. For all the metrics, we report the obtained $p$-value, and consider the result significant if $p < 0.01$.

\subsubsection{Results}

\begin{table}[t]
    \centering
    \caption{The results of the online evaluation. The bold font indicates better results (whether lower or higher). The star indicates a statistically significant result ($\mathbf{p < 0.01}$).}
    \vspace{-0.2cm}
    \begin{tabular}{ccccc}
        \toprule
        \textbf{Metric} & \textbf{Heuristics} & \textbf{CatBoost} & $\mathbf{p}$-\textbf{value}\\\midrule
        \textbf{Explicit select} {\small (sessions)} & 0.247 & \textbf{0.292} & 0.018 \\
        \textbf{Typed select} {\small (sessions)} & 0.416 & \textbf{0.346*} & < 0.001 \\
        \textbf{Typing actions} {\small (symbols)} & 2.073 & \textbf{1.832*} & 0.008 \\
        \textbf{Prefix length} {\small (symbols)} & 2.628 & \textbf{2.275*} & < 0.001 \\
        \textbf{Manual start} {\small (sessions)} & 0.047 & \textbf{0.079*} & < 0.001 \\
        \bottomrule
    \end{tabular}
    \vspace{-0.5cm}
    \label{tab:evalOnline}
\end{table}

Table~\ref{tab:evalOnline} summarizes the results of the online evaluation comparing the models. Once again, it can be seen that all the metrics indicate the better performance of the CatBoost-based model trained on users' logs. The fraction of the sessions that ended with explicit selection increased, however, this result did not pass the statistical significance test. At the same time, the fraction of sessions that ended with the manual typing of the entire token decreased significantly. Next, we can see that the number of the necessary characters to type also decreased. The overall average number of typing actions decreased from 2.073 to 1.832, and the average length of prefix at the moment of explicit selection decreased from 2.628 to 2.275. These two metrics indicate that the model actually makes it easier for the users to input a token. Finally, the fraction of sessions that started manually increased significantly, from 0.047 to 0.079, which might indicate the users' interest in obtaining the results of code completion.

\subsection{Overall Performance}

Lastly, it is necessary to discuss whether the obtained model complies with the limitations described in Section~\ref{sec:model}, since this also directly impacts the comfort of the end-users.

The size of the trained model was 366 KB, which comfortably fits into the desired range. As for the latency, the A/B test showed that when moving from heuristics to CatBoost, it increased from 92.3 ms to 119.3 ms ($p$ < 0.01). This corresponds to the model adding less than 30 ms for inference, thus also remaining in the comfortable range, virtually unnoticeable for the user.

Overall, it can be seen that the Decision Tree based model trained on real usage logs demonstrated its superiority over the default heuristics-based ranking in all tested settings, while remaining small enough and fast enough to be used in production. Even more importantly, the proposed approach allows us to collect more data, design more features, and evaluate more models continuously, making sure that they always remain relevant. 
\section{Discussion \& Open Challenges}
\label{sec:discussion}

\subsection{Code Completion}

\subsubsection{Model.} We show that leveraging real-world structured IDE usage logs is beneficial for both training ML models (similar to Aye et al.~\cite{Aye2020real}) and evaluating their performance (similar to Proksch et al.~\cite{Proksch2016eval}). However, we only use anonymous pre-extracted features instead of the full record of development history or edit context.

Svyatkovskiy et al.~\cite{svyatkovskiy2021fast} argue in favor of using end-to-end neural networks, and point out that feature-based models...

\begin{itemize}
    \item[---] \textit{...depend on hard-coded features, thus missing the opportunity to learn richer features directly from the data.} Although this is true, in our case, it becomes an advantage --- we use this as a \textbf{safeguard mechanism} to control what data gets collected, and thus prevent gathering sensitive information.
    \item[---] \textit{...introduce difficulties in manually designing and extracting relevant features that cover as many cases as possible.} This is also true, but every major language has a dedicated team to support it, and also, \textbf{organization-wide tooling and infrastructure} allow us to automate the experiments and lead to continuous improvements of the overall system.
    \item[---] \textit{...learn about individual APIs and cannot generalize to unseen ones.} Our approach does not have this problem, as for providing the list of suggestions, it relies on \textbf{project-wide static analysis and context-specific feature extraction}, and not on a global vocabulary.
\end{itemize}  
 
As far as a specific model goes, we decided on using \textit{Decision Tree} based models, specifically, CatBoost. This choice is dictated mainly by the strict limitations that our task imposes. Such models are simple, interpretable, and have great runtime performance: small size and low latency. Additionally, they turned out to be simple to work with: they provide fast training, great production-grade tooling, and, importantly, they are easily converted into an intermediate representation that allows to re-use them between different ecosystems (\textit{e.g.}, Python and JVM).

At the same time, it must be noted that the practical nature of our task does not allow us to claim state-of-the-art results or even compare directly with many models from the literature. Our goal in this continuing research is precisely to find models that would improve the user experience while being lightweight and easy to use in an actual IDE on a consumer-grade device.

\subsubsection{Data.} An interesting open question when using logs of completion sessions is what exactly to consider positive examples and what to consider negative examples. A default idea that we used consists of taking all the sessions that ended with a certain positive outcome (\textit{i.e.,} \textit{Explicit select} or \textit{Typed select}), using their selections as positive examples and anything that was not selected as negative examples. However, this leaves out all the ``fully negative'' cases, \textit{i.e.}, \textit{Explicit cancel} and \textit{Typed cancel} that can still provide negative examples. Also, it can be noticed that we use \textit{Typed select} as a source of positive examples, however, in our A/B tests (see Section~\ref{sec:online}), we try to \textit{lower} the ratio of users who type the token themselves. This can also be considered when selecting positive and negative examples. Even more granularly, we can say that since our goal is to make the users type less (hence, using \textit{typing actions} as a metric), not all positive examples are of the same value. It is possible to introduce a weight to the examples based on their length, to facilitate the model to improve the raking of the longer suggestions. For example, users may fully type very simple keywords like \texttt{for} or \texttt{def} almost immediately, thus making them positive examples of the \textit{Typed select} class, however, they might not be what we want the model to learn.

\subsubsection{API usage.} Our analysis of the user logs shows that code completion demonstrates the worst results when it is necessary to suggest API calls --- both internal and external. Oftentimes, the information about the local context is not enough, be that file, module, or even the entire project, --- instead, one may look at how a particular API is used in other projects. A promising direction is using a corpus of open-source code to learn similar context and incorporating this information into the ranking.

\subsection{Logs and Infrastructure}

Importantly, the goal of our research and this paper does not lie only in the area of code completion. Rather, we want to emphasise the importance of structured user logs, and how a pipeline for their collection can be used to both develop features and evaluate them.

Besides code completion, the same described infrastructure can be used to improve other features of the IDE: refactoring recommendation~\cite{kurbatova2020recommendation}, code smell removal~\cite{de2020casper, lambiase2020just}, and others. Collecting user logs allows us to not make any assumptions about their behavior, or at least move away from them, and instead continuously take into account their feedback. This pipeline can be used to bridge academic results with practical applications. If researchers develop a new model for ranking code completion, we can collect its output as a feature, use it when training our models, and see in practice whether it improves the user experience.

Overall, our experimental results and the experience of running such a system in production for more than a year demonstrate that it is a valuable tool for improving the IDE. We were able to run the described pipeline for many prominent languages --- Java, Python, Kotlin, JavaScript, TypeScript, Ruby, Go, and others, --- with the infrastructure of the experiments being largely reused. The default completion ranking models in many IntelliJ-based IDEs right now are the ones trained on the user logs. Moreover, the process continues to this day, with new features being evaluated and new models being tested.
\section{Related Work}
\label{sec:related}

A lot of research has been dedicated to improving code completion using machine learning methods. As we already mentioned, a principal distinction between different works is whether they are trained on synthetic data or the real user data~\cite{hellendoorn2019code}. In this section, let us describe several key studies.

Bruch et al.~\cite{Bruch2009} proposed to use three simple ML approaches to improve code completion, namely, a straightforward frequency-based system, an association rule mining, and a modification of the $k$ Nearest Neighbors approach. The authors focused on the API recommendations and used synthetic data: an existing code base, where some method calls were removed to simulate completion queries. Of the three tested models, the latter showed the best results in terms of the F1-measure.

Proksch et al.~\cite{Proksch2015inteli} proposed to use Bayesian networks instead. Their work is also limited to API calls and deals with synthetic data, however, the authors suggested that a proper evaluation of a code completion system must rely not only on the quality analysis, but also on the performance analysis. For this reason, the authors also took into account the size of the developed model and the inference speed, and tested it on queries of different size.

More recently, Svyatkovskiy et al.~\cite{svyatkovskiy2021fast} evaluated end-to-end neural networks instead of feature engineering based approaches. The authors also formulated the task as a learning to rank problem and leveraged static analysis as a candidate provider. Similarly, the authors also took into account the importance of model size and inference speed, and thus evaluated them too. In a thorough comparison, the authors evaluated different token encoders and different context encoders. At the same time, in this work, only the API recommendations are addressed, and synthetic data is used (the source code of the most-starred Python projects). The approaches proposed by Svyatkovskiy et al. have been used as a base for code completion solutions for the Visual Studio Code IDE~\cite{svyatkovskiy2019pythia, svyatkovskiy2020intellicode}.

The importance of using real-world user data was carefully studied in the seminal paper by Hellendoorn et al.~\cite{hellendoorn2019code}. The authors collected a dataset of 15,000 real completions conducted by 66 users and compared them with existing synthetic benchmarks. Not only did the authors find that some state-of-the-art techniques demonstrate significantly worse performance on the real data, they also show that some features of the real-life code completion usage are invisible in a synthetic setting: specifically, the users spending the most time on the infrequent tokens that the models recommend with even worse quality.

Similarly, Proksch et al.~\cite{Proksch2016eval} evaluated different approaches on the real-world data of 7,157 queries and also showed that synthetic evaluations provide unrealistic numbers when compared to the ground truth. However, these works used real user data only for offline evaluation, not for actually training better models on them --- probably, since it is difficult to collect a necessary amount of data in the research setting.

Finally, Aye et al.~\cite{Aye2020real} did train a model on the real-world data, and also carried out an A/B test to evaluate it. More specifically, the authors studied the Hack dialect of PHP and trained end-to-end neural network models on three different datasets: (1) \textit{Baseline} --- synthetic static codebase of nearly one million source files, (2) \textit{Autocompletion} --- real accepted completions inside the IDE, and \linebreak (3) \textit{Edit} --- code edits logged during file-save operations. Then, among other experiments, the authors conducted two live A/B tests, comparing models trained on two latter datasets to the model trained on the \textit{Baseline} dataset. The results showed that the model trained on the \textit{Autocompletion} dataset performed the best, indicating the usefulness of the user behavior logs. However, it is important to note that this work was carried out inside Facebook, which is why the authors could collect personal information (specific code snippets in logs), so, while their datasets are undoubtedly large, they are not limitless and may not generalize well to all users.

Overall, it can be seen that our work compliments the existing ones: it uses the real-world usage logs to both train and evaluate models, develops an approach that takes into account the users' privacy, and implements a pipeline for the continuous gathering of data and evaluation of models using the IDE release cycle.

\vspace{-0.2cm}
\section{Threats to Validity}
\label{sec:threats}

The industrial nature of the solved problem and the orientation of the solution towards production impose certain limitations on our research. Several important threats to validity can be highlighted.

\textbf{User bias.} Our approach relies on the logs of users, training the model on their decisions, with the goal of improving the experience of all the subsequent users of the IDE. However, the users who participate in the Early Access Program and who agree to send the logs might not be a perfectly representative sample of all the users. These users are in general more active and can be more aware of various IDE features. What is more, their activity and their work in the IDE may also differ from that of an average user. At the same time, a large sample of thousands of users (and tens of thousands of completion sessions) allow us to gather data from different domains of software engineering and different levels of activity.

\textbf{State-of-the-art.} As mentioned in Section~\ref{sec:discussion}, the positioning of our work does not allow us to claim state-of-the-art results or directly compare with them, making it possible for us to have missed a specific work or model architectures that would perform better. However, the main goal of this work was to present a pipeline for the \textit{continuous} improvement of IDE tooling, making it possible to compare the necessary models in the near future.

\textbf{Features.} The same argument can be applied to the list of the calculated features. Privacy limitations lead us towards the manual development of features, making it possible for our current set to not be optimal. Once again, the developed infrastructure allows us to evaluate new features constantly and test any new developments in a continuous cycle.

While these threats are important to note, we believe that they do not invalidate the usefulness of the proposed pipeline and the results of the carried out evaluations.

\vspace{-0.2cm}
\section{Conclusion \& Future Work}
\label{sec:conclusion}

In this work, we presented a pipeline for collecting anonymous logs from users to train a model for ranking code completion suggestions. We designed a set of features that are calculated on the user's machine and are then anonymously collected to the server without gathering sensitive personal information or the code itself. Based on these features, we trained a CatBoost model for ranking completion candidates and evaluated it in two settings. The offline evaluation on the held-out set of user data showed that the Recall at K of the ranking increases when using the model. The online evaluation consisted in an A/B test between the trained model and the default heuristics-based ranking, and showed that the fraction of sessions that ended with the explicit selection of the token increased, while the average number of the required typing actions decreased.

An important aspect of the proposed idea is that it can be used continuously, gathering new data and evaluating new models in a constant cycle. This is implemented in the Early Access Program (EAP) of IntelliJ-based IDEs: users may agree to anonymously send their logs to the centralized server in exchange for being able to test new features. In each EAP release, before the main release of the IDE, the experiments are being conducted: users are divided into groups, some of which get the current best performing models, while some get new ML-based models that are then compared in a live A/B test. At the same time, logs are also being collected from other users to train future models. This pipeline has been in production since the end of 2020, and demonstrated its usefulness for designing new features and discovering better models.

The proposed pipeline may be used for other tasks beyond code completion. Various anonymous features may be collected and used to fine-tune different parts of the IDE: suggesting refactorings, fixing bugs, etc. We hope to see more research that relates to collecting structured usage logs. Such a setting also puts research-based approaches into a real-world environment, where the tested models have to be quick and lightweight. 
In future work, we plan to continue our experiments with code completion models and develop new features that would allow them to perform better. In particular, we are working on improving code completion for API calls using the information about similar contexts in other projects. We would also like to broaden our scope and employ such machine learning models to other aspects of software development in the IDE.

\bibliographystyle{ACM-Reference-Format}
\balance
\bibliography{literature}

\end{document}